\documentclass{emulateapj}
\def\DiskFit{{\it DiskFit}}
\def\ie{{\it i.e.}}
\def\etal{{\it et al.}}

\def\etc{{\it etc.}}

\long\def\Ignore#1{\relax}

\slugcomment{Posted \today}

\shorttitle{\DiskFit}
\shortauthors{Sellwood \& Spekkens}

\begin{document}

\title{\DiskFit: a code to fit simple non-axisymmetric galaxy models
  \\ either to photometric images or to kinematic maps}

\author{J. A. Sellwood}
\affil{Department of Physics and Astronomy, Rutgers University, \\
    136 Frelinghuysen Road, Piscataway, NJ 08854, USA}
\email{sellwood@physics.rutgers.edu}

\and

\author{Kristine Spekkens}
\affil{Department of Physics, Royal Military College of Canada, \\
  PO Box 17000, Station Forces Kingston, ON K7K 7B4, Canada}
\email{Kristine.Spekkens@rmc.ca}

\begin{abstract}
This posting announces public availability of version 1.2 of the
\DiskFit\ software package developed by the authors, which may be used
to fit simple non-axisymmetric models either to images or to velocity
fields of disk galaxies.  Here we give an outline of the capability of
the code and provide the link to downloading executables, the source
code, and a comprehensive on-line manual.  We argue that in important
respects the code is superior to {\it rotcur\/} for fitting kinematic
maps and to {\it galfit\/} for fitting multi-component models to
photometric images.
\end{abstract}


\section{Introduction}
\label{intro}
The code \DiskFit\ may be used to fit simple non-axisymmetric models
either to images or to velocity fields of disk galaxies.  If the fit
is successful, \DiskFit\ provides quantitative estimates of the
non-circular flow speeds and an estimate of the mean circular speed
when run on velocity fields \citep{SS07, SZS10}, and fractions of the
galaxy light in a bar, disk, and bulge when run on images
\citep{Rees07, Kuzi12}.

The kinematic branch of the code differs fundamentally from the
frequently-used {\it rotcur\/} algorithm \citep{Bege87} since the
minimization fits a global model to the complete map, rather than
separate tilted rings, and it is superior to {\it reswri\/}
\citep{Scho97} because it does not employ the epicyclic approximation
to fit departures from circular motion.  The photometric branch of the
code also differs fundamentally from popular algorithms such as {\it
  galfit\/} \citep{Peng10} in that it fits non-parametric disk and bar
light profiles rather than specified functional forms.  Furthermore,
it is superior to all three algorithms because it is capable of
providing statistically valid, but realistic, estimates of the
uncertainties in the fit.  \citet{Kuzi12} illustrate the functionality
of \DiskFit\ on high-quality kinematic and photometric data for the
nearby galaxy NGC~6503.

A single code is provided to fit both photometric images and kinematic
maps because, for both applications, \DiskFit\ employs the same basic
minimization algorithm originally described in the Appendix of
\citet{BS03}.  The first applications \citep{BS03} were to fit
axisymmetric models; the extension to non-circular flows was described
by \citet{SS07}\footnote{\DiskFit\ is also an extension of the
  publicly-available {\it velfit} 2.0.} and by \citet{SZS10}, while
\citet{Rees07} extended the code to include barred models when fitting
photometric images.

Note that \DiskFit\ does not fit photometry and kinematics
simultaneously; the same code simply fits either type of data
depending on the user’s choice of inputs.  Of course, if the user
makes separate fits to both types of data from the same galaxy, the
fitted values will likely differ.

\section{Capabilities of the code}
\DiskFit\ minimizes a $\chi^2$ estimate of the differences between a
projected model and the data.  The data can be either a 2D velocity
map derived from Doppler shifts of spectral lines obtained using an
IFU in the optical or aperture synthesis in the radio, or a
photometric image.  The user can supply a map of uncertainties in
the data and a mask image to indicate only good pixels to be fitted.

\subsection{Fitting an axisymmetric model}
Aside from an optional simple warp in the outer parts, the model
presented to the data is a flat disk with inclination, $i$, and
position angle, $\phi_d$, that are assumed to be the same at all
radii.  Furthermore, the position of the center, $(x_c,y_c)$ and, for
kinematic fits only, the systemic velocity, $v_{\rm sys}$, are
parameters fitted to the entire 2D data set.  A simple axisymmetric
model will therefore fit any or all these parameters to determine
global estimates that best fit the data.

In addition, \DiskFit\ estimates either the circular speed, for
kinematic data, or the mean intensity for a photometric image, at a
set of radii specified by the user.  Model values at data points that
lie between the specified radii are computed by linear interpolation.
It is important to note that this implies the model is simply a
tabulated set of values over a range of radii and has {\it no
  pre-specified functional profile}, such as an exponential disk, \etc

\newpage
\subsection{Uncertainties}
Uncertainties in the parameters, and in the intensity or circular
speed at each radius, are estimated by a bootstrap method.  The
residuals from a simple model are generally correlated at neighboring
pixels, because the model ignores spirals and other sources of
correlated turbulence.  The bootstrap algorithms employed attempt to
preserve these correlated residuals \citep[see][for a fuller
  discussion]{SS07, SZS10}, which lead to larger and more realistic
estimates of the uncertainties in the model.

\subsection{Non-axisymmetric models}
The most powerful aspect of \DiskFit\ is that it can include simple
non-axisymmetric features into the model and fit for their parameters.
The most useful capability is to fit for a bar, which is a
bi-symmetric distortion having a fixed position angle that is, in
general, not aligned with, or perpendicular to, the major axis of
projection.  \DiskFit\ allows for an underlying axisymmetric model on
which a non-axisymmetric feature having a fixed position angle in the
disk plane is superposed, and returns an estimate of the angle of its
principal axis to the disk major axis.  A bar that is almost aligned
with the major or minor axis of projection may require that the fit is
smoothed (see \S2.7), but the bar cannot be separated from the disk by
this algorithm when the alignment is exact; note that in such a case,
an axisymmetric fit will be no worse than that obtained by other
algorithms.

For kinematic fits, the non-circular flows have two \hbox{$m$-fold}
symmetric components ($m=2$ for a bar): a radial part that is the mean
flow away from and towards the model center, and an azimuthal part
that is the departure above and below the mean streaming speed.  Each
component varies in azimuth in the disk plane as a $\cos(m\theta)$ or
a $\sin(m\theta)$ function, respectively, with zero phase on the bar
major axis.  These additional velocities are fitted at the same radii
as those used to tabulate the circular speed, although the user can
specify that the distortion has a smaller radial extent than the
entire disk.  \DiskFit\ does not impose any relation between the
radial and azimuthal velocity distortions, which can be arbitrarily
large compared with the mean circular speed -- \ie\ it is not
restricted to a small amplitude distortion.  If the distortions turn
out to be small, \cite{SZS10} give formulae that can relate the fitted
velocity distortions to the ellipticity of the potential.

For photometric fits, the bar represents a light component that
increases the fitted intensity above the axisymmetric mean along the
bar major-axis, with a corresponding reduction along the bar minor
axis.  The bar light profile is again tabulated at the same radii as
the mean axisymmetric light profile.

In principle, \DiskFit\ can fit for distortions having other
rotational symmetries, such as $m=1$ (lopsided) or $m=3$ (trefoil)
distortions, although they could not be spiral in form as the
algorithm restricts the non-axisymmetric component to having a fixed
position angle in the disk plane at all radii.

Less usefully, \DiskFit\ can also fit for axisymmetric radial flows.
However, radial flow velocities would need to be unrealistically large
-- at least a few percent of the circular speed -- to be detectable.
Axisymmetric flow speeds of this magnitude would indicate the galaxy
is in a transitional state and that extensive rearrangement of the
mass distribution is taking place on a dynamical time-scale.

\subsection{Spiral distortions}
The largest residuals in fitted models generally arise from spiral
arms, which are non-circular flows in kinematic maps and coherent
features in photometric images.  \DiskFit\ does not attempt to fit
these distortions, and merely treats them as sources of error that
are allowed for in the bootstraps.

The reason is that these features are hard to model.  Unlike bars,
which are strong, clearly bisymmetric, and long-lived, mild spiral
distortions are transient and probably result from multiple,
superposed modes having different pattern speeds, and rotational
symmetries.

\subsection{Warp fitting}
\DiskFit\ allows the model to be warped in a simple, parametric
manner.  The code assumes that the line of nodes of the warp is at a
fixed position angle, the warp begins at a certain radius, and
increases in amplitude as a quadratic function of radius to some
maximum amplitude at the last measured point.  Since the kinematic
signature of a warp closely resembles that of an in-plane bar,
\DiskFit\ will not allow the user to select both options in the same
fit.

\subsection{Bulge fitting}
Photometric images can be fitted with a disk, bar and bulge model if
desired.  \DiskFit\ makes the (highly questionable) assumptions that
the bulge is both axisymmetric and symmetric about the disk mid-plane,
and has a flattening that is constant with radius.  It also assumes
the parametric form of a S\'ersic profile for the bulge, and will fit,
if desired for the S\'ersic index, $n$, effective radius, $R_e$,
central intensity, $I_0$, and flattening $\epsilon_b$.  A very high
spatial resolution image is generally required to fit for all these
parameters, and it is usually safer to hold at least $n$ fixed at some
reasonable value.

The user of this capability should bear in mind that the fitted values
provided by the code are meaningful only if the above listed
assumptions about the bulge light profile are valid for his/her data.

\subsection{Seeing corrections}
If the user requests, \DiskFit\ will blur the model, by convolving it
with a point spread function, before comparing it with the data, which
can be done for either photometric or kinematic fits.  The blurring
function is a Gaussian of specified width; note that the FWHM cannot
be greater than 3 pixels.  The code to compute these seeing
corrections had a bug in version 1.1, which has been fixed in the
present release.

\subsection{Smoothing penalties}
In general, \DiskFit\ places no restrictions on the tabulated values
of radial variation of the light profile, rotation curve, bar
distortion amplitudes, \etc \ We note that \DiskFit\ has an option to
apply a smoothing penalty to the radial variation of these tabulated
functions, if desired.  Since the smoothing penalty will affect the
fitted values, it should never be large, and no smoothing is
recommended in most cases.  However, \citet{SZS10} found that when
fitting for the flow velocities of a bar that was inclined by just a
small angle to the projected major axis, the velocity distortions
became absurdly large and variable, and some smoothing was necessary
to obtain meaningful fits.

\section{Obtaining \DiskFit}
The code is available from {\tt http://www.physics\break
  .rutgers.edu/$\sim$spekkens/diskfit}.  This website includes
links to a comprehensive manual, giving full details of the procedure
to use the code, data requirements, and illustrative examples,
software update history, as well as executables and the source code.
Versions 1.0 and 1.1 were released previously, in September 2012 and
May 2013 respectively, and this posting is to announce version 1.2.
The improvements at this version are that arrays are dimensioned
dynamically, so that there are no software limits to the size of the
dataset that can be fitted, and several bugs have been fixed.

The authors encourage feedback from users, and will make every effort
to correct bugs and inconsistencies.  Requests for additional
capabilities will be considered and may be provided in future
releases, but the authors cannot undertake to meet every possible
request.

\section{Copyright and license issues}
\DiskFit\ is free software and comes with ABSOLUTELY NO WARRANTY.  It
is distributed under the GNU General Public
License.\footnote{\label{note.GNUl}see
  http://www.gnu.org/copyleft/gpl.html} This implies that the software
may be freely copied and distributed.  It may also be modified as
desired, and the modified versions distributed as long as any changes
made to the original code are indicated prominently and the original
copyright and no-warranty notices are left intact.  Please read the
General Public License for more details.

Note that the authors retain the copyright to the code and
documentation.  Those publishing papers that use the code are
requested to acknowledge this arXiv posting as the source.


\begin{thebibliography}{}
\def\PhD{Ph.D.\ thesis}

\bibitem[Barnes \& Sellwood(2003)]{BS03}
Barnes, E. I. \& Sellwood, J. A. 2003, \aj, {\bf 125}, 1164 

\bibitem[Begeman(1987)]{Bege87}
Begeman, K. G. 1987, \PhD, University of Groningen

\bibitem[Kuzio de Naray \etal\ (2012)]{Kuzi12}
Kuzio de Naray, R., Arsenault, C. A., Spekkens, K., Sellwood, J. A., McDonald, M., Simon, J. \& Teuben, P. 2012, \mnras, {\bf 427}, 2523

\bibitem[Peng \etal\ (2010)]{Peng10}
Peng, C. Y., Ho, L. C., Impey, C. D. \& Rix, H-W. 2010, \aj, {\bf 139}, 2079

\bibitem[Reese \etal\ (2007)]{Rees07}
Reese, A., Williams, T. B., Sellwood, J. A., Barnes, E. I. \& Powell, B. A. 2007, \aj, {\bf 133}, 2846

\bibitem[Schoenmakers \etal\ (1997)]{Scho97}
Schoenmakers, R. H. M., Franx, M., \& de Zeeuw, P. T. 1997, \mnras, {\bf 292}, 349 

\bibitem[Sellwood \& Z\'anmar S\'anchez (2010)]{SZS10}
Sellwood, J. A. \& Z\'anmar S\'anchez, R. 2010, \mnras, {\bf 404}, 1733

\bibitem[Spekkens \& Sellwood (2007)]{SS07}
Spekkens, K. \& Sellwood, J. A. 2007, \apj, {\bf 664}, 204

\end{thebibliography}
\end{document}